\UseRawInputEncoding
\documentclass[journal]{IEEEtran}
\usepackage{xcolor,soul,framed} %,caption
\usepackage{graphicx}
\usepackage{amsmath}
\usepackage{array}
\usepackage{mdwmath}
\usepackage{mdwtab}
\usepackage{eqparbox}
\usepackage{url}
\usepackage{lineno,hyperref}
\usepackage{amsmath,amssymb}
\usepackage{booktabs}
\usepackage{makecell}
\usepackage{multirow}
\usepackage{multirow}
\usepackage{tabu,bm}
\usepackage{color}
\usepackage{lipsum,cuted}
\usepackage{stfloats}
\usepackage{cite}
\usepackage{enumitem}
\usepackage{diagbox}
\usepackage{subcaption}
\usepackage{tabularx}
\usepackage{array}
\usepackage{geometry}
\usepackage{setspace}
\newcolumntype{L}[1]{>{\raggedright\arraybackslash}p{#1}}
	
\title{From Specialist to Large Models: A Paradigm Evolution Towards Semantic-Aware MIMO}

\author{Keke Ying, Zhen Gao, Tingting Yang, Jianhua Zhang, Xiang Cheng, Tony Q.S. Quek, and H. Vincent Poor
\thanks{Keke Ying and Zhen Gao (\textit{corresponding author}) are with Beijing Institute of Technology, China; Tingting Yang is with Peng Cheng Laboratory, China; Jianhua Zhang is with Beijing University of Posts and Telecommunications, China; Xiang Cheng is with Peking University, China; Tony Q. S. Quek is with Singapore University of Technology and Design, Singapore; H. Vincent Poor is with Princeton University, USA.}
}

\begin{document}

\maketitle

\begin{abstract}
The sixth generation (6G) network is expected to deploy larger multiple-input multiple-output (MIMO) arrays to support massive connectivity, which will increase overhead and latency at the physical layer. Meanwhile, emerging 6G demands such as immersive communications and environmental sensing pose challenges to traditional signal processing. To address these issues, we propose the ``semantic-aware MIMO'' paradigm, which leverages specialist models and large models to perceive, utilize, and fuse the inherent semantics of channels and sources for improved performance. Moreover, for representative MIMO physical-layer tasks, e.g., random access activity detection, channel feedback, and precoding, we design specialist models that exploit channel and source semantics for better performance. Additionally, in view of the more diversified functions of 6G MIMO, we further explore large models as a scalable solution for multi-task semantic-aware MIMO and review recent advances along with their advantages and limitations. Finally, we discuss the challenges, insights, and prospects of the evolution of specialist models and large models empowered semantic-aware MIMO paradigms.
\end{abstract}

\section{Introduction}
Massive multiple-input multiple-output (MIMO) technology is expected to be a pivotal technology for sixth generation (6G) wireless systems. On the base station (BS) side, supporting immersive communication scenarios such as mixed reality necessitates larger antenna arrays to achieve improved spatial diversity and multiplexing capabilities. On the user equipment (UE) side, the dramatic growth of device density in the Internet of Things calls for more advanced multi-user access schemes to handle massive connectivity. Besides, MIMO channels inherently contain rich information about the wireless propagation environment \cite{qzj-wc}, efficiently exploiting such structure for communication or sensing remains challenging.

Standard MIMO theory faces challenges due to the vast MIMO dimensions and signal modal diversity. For instance, conventional methods for MIMO channel state information (CSI) acquisition rely on signal detection and estimation theory, aiming to acquire complete CSI or parametrized information such as multipath delays, angles, and Doppler shifts. As BS antennas or transmission environment complexity increases, these methods demand excessive pilot overhead. Regarding source transmission tasks, traditional schemes aim to design source and channel codec schemes that approach channel capacity within the syntax communication framework of information theory \cite{qzj-wc}. For data-intensive sources, such as images and videos, significant data transmission overhead is required, and a cliff effect occurs where performance severely degrades when the quality of channel falls below a certain threshold.

Recently, semantic communication has gained attention \cite{qzj-wc} as it focuses on efficiently and accurately transmitting the meaning of source information, unlike traditional syntactic communication, which aims to accurately recover transmitted symbols.  By exploring intrinsic meanings or task-related semantic information, the transmission resource requirements can be greatly reduced. 
This encourages the design of MIMO systems aware of source and channel semantic features to solve physical-layer tasks, potentially reducing transmission overhead or enhancing performance metrics in adverse transmission conditions.
Achieving this involves leveraging the semantic encoding capabilities of deep neural networks (DNNs), which can be learned from large amounts of data. In MIMO systems, channel and source semantics can be transformed into feature representations using DNNs. For example, the Transformer framework \cite{wy-wc} allows conversion of source and channel modalities into feature sequences, enabling the Transformer to extract task-specific semantics for various tasks.

More recently, Inspired by the success of large language models (LLMs) in natural language processing, researchers are exploring how to integrate specialized models into a universal communication large model (LM) framework. This approach seeks to align diverse physical-layer transmission tasks at a unified token space, allowing a single LM to manage multiple tasks efficiently. For example, the mixture of experts (MoE) method offers an efficient solution by activating only a subset of expert subnetworks within the LM during inference. This enables scalable and adaptive task management without significantly increasing computational costs, showcasing potential for handling various tasks.

In the remaining sections of this article, we will first discuss several semantic-aware MIMO communication cases based on specialist models, including user activity detection (UAD), CSI feedback, and multi-user precoding, to reveal how they utilize deep learning for the perception, utilization, and fusion of semantic information in MIMO systems. Next, we will further investigate and compare existing research on MIMO physical-layer tasks based on general LMs. Following that, we will discuss challenges and future directions. The final section summarizes the entire article.

\section{Specialist Models for Semantic-Aware MIMO}
\begin{figure}[t]
	%\vspace{-5mm}
	\centering
	\captionsetup{font={footnotesize}, name={Fig.},labelsep=period}
	\includegraphics[width=\linewidth]{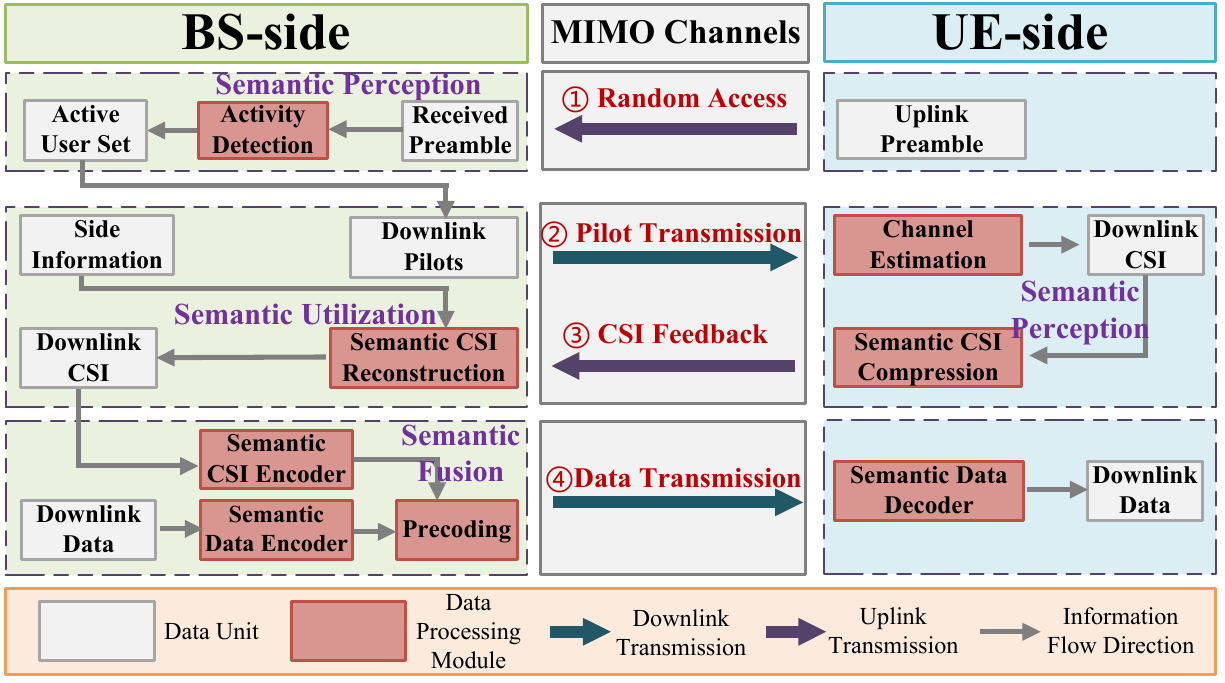}
	\caption{System architecture of a semantic communication framework incorporating key functionalities including activity detection, CSI feedback, and downlink precoding. The communication process involves three semantic stages: (i) semantic perception, which extracts local features such as user activity and channel characteristics during activity detection and semantic CSI compression; (ii) semantic utilization, where side information is leveraged to enhance CSI reconstruction quality; and (iii) semantic fusion, involving the fusion of semantic data and downlink CSI for precoding design.}
	\label{fig1}
	\vspace{-4mm}
\end{figure}

In this section, we examine semantic-aware specialist models for physical-layer MIMO tasks, including uplink random access, CSI feedback, and precoding. These tasks highlight uplink UE-side semantics, propagation-environment semantics, and downlink BS-side semantics, providing a comprehensive view of the wireless communication pipeline. As illustrated in Fig. \ref{fig1}, the process starts when users request initial connection with the BS via random access. The BS must then identify active users from the superimposed preamble sequence. Subsequently, before downlink transmission, obtaining high-dimensional CSI through the feedback link is essential for precoding, especially for systems without channel reciprocity. Finally, the BS performs precoding for high-quality multi-user transmission. Focusing on these tasks, we review existing research and present semantic-aware designs to enhance performance. The Transformer architecture was selected as the backbone of our semantic-aware MIMO framework. Its self-attention mechanism effectively captures global dependencies, while cross-attention flexibly models correlations across heterogeneous domains or data sources. Additionally, the Transformer architecture provides efficient parallel computation, scalability in model size, and seamless integration with the large-model paradigm.

\begin{figure*}[b!]
	\vspace*{-5mm}
	\begin{center}
		\hspace*{-4mm}\subfloat[]{
			\label{fig.aud_a}
			\includegraphics[width=.65\textwidth,keepaspectratio]{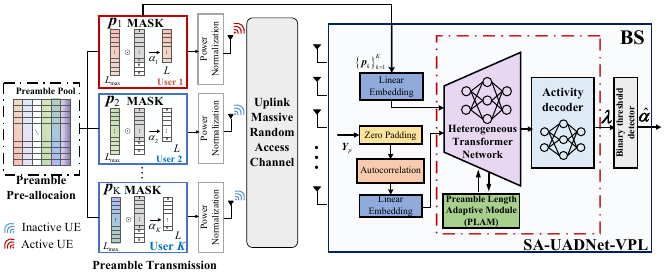}
		}
		\hspace*{-2mm}\subfloat[]{
			\label{fig.aud_b}
			\includegraphics[width=.35\textwidth,keepaspectratio]{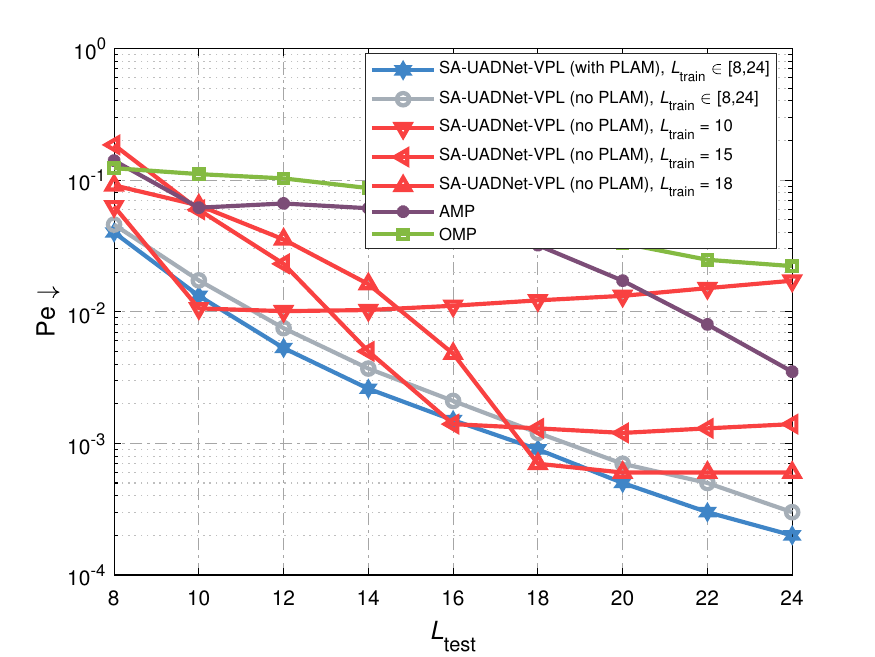}
		}
	\end{center}
	\vspace*{-5mm}
	\captionsetup{font={footnotesize}, name={Fig.},labelsep=period}
	\caption{(a)~System diagram of semantic-aware massive random access with variable preamble length transmission; and (b)~Performance of $P_e$ relative to testing preamble lengths $L_{\text{test}}$ in various UAD schemes. \textbf{Notes:} During simulation, we randomly generated QPSK preambles of maximum length $L_{\text{max}}=28$, with $K=128$ potential UEs and a BS equipped with 64 antennas. The activity of each UE follows a Bernoulli distribution with an active probability of 0.1, and the channels between UEs and the BS follow independent and identically distributed Rayleigh fading. }
	\label{fig.aud}
	\vspace*{-4mm}
\end{figure*}
\subsection{UAD for Random Access}
\subsubsection{Related Work}
With the growing density of cellular network devices, identifying an active user set (AUS) from massive numbers of UEs is a crucial initial step for 
establishing efficient communication link.  Grant-free non-orthogonal access effectively reduces massive random access delays. In this method, shown in the upper part of Fig. \ref{fig1}, each user is assigned a specific non-orthogonal preamble, and the BS needs to identify AUS from the received superimposed preambles of all potential users. This activity information extraction can be seen as a form of semantic perception. 

Existing methods like compressed sensing (CS) \cite{kml-tsp} and covariance-based maximum likelihood detection \cite{TransAUD} address this by leveraging sparsity or building likelihood models. CS focuses on reconstructing sparse channel matrices rather than directly detecting active users, increasing preamble overhead. The covariance-based approach improves detection performance at the cost of higher complexity. \cite{TransAUD} further introduces a heterogeneous Transformer to learn correlations between received signal covariance and preambles, boosting performance and lowering complexity. Yet, data-driven methods struggle to generalize across different preamble lengths within a single model, preventing dynamic adaptation to traffic and wasting resources. This calls for adaptable detection networks that can efficiently capture user activity semantics across variable preamble lengths.

\subsubsection{Semantic-Aware UAD with Variable Preamble Length}
We propose a semantic-aware UAD network with variable preamble length (SA-UADNet-VPL) that directly perceives user activity semantics instead of reconstructing the sparse channel matrix, enabling a single model to handle different preamble lengths. As shown in Fig. \ref{fig.aud}(a), in an uplink access system, each of the $K$ potential users is pre-assigned a unique preamble sequence $\boldsymbol{p}_k$ of length $L_{\text{max}}$. When some users become active, they transmit their respective preamble sequences. The BS receives the superimposed preamble and determines the AUS.

Extending \cite{TransAUD} to variable preamble lengths, we leverage a Transformer encoder to extract correlations. Specifically, we project user preambles $\{\bm{p}_k\}_{k=1}^{K}$ and the autocorrelation of the zero-padded received signal $\bm{Y}_p$ into same-dimensional features using two linear layers. Given their distinct semantics, we use a heterogeneous Transformer with separate weights for $\bm{Y}_p$ and $\{\bm{p}_k\}_{k=1}^{K}$ in multi-head attention and feed-forward layers. In the final output layer, the activity decoder calculates the correlation between the feature sequences corresponding to $\bm{Y}_p$ and the multi-user preambles $\{\bm{p}_k\}_{k=1}^{K}$ to estimate the user activity $\bm{\lambda}$. The proposed scheme employs the categorical cross-entropy of user activity as its objective function, allowing the receiver to directly capture activity semantics from the received preamble.

To improve scalability, we introduce a preamble-length adaptive mechanism. The transmitter sets a maximum preamble length $L_{\text{max}}$; when the actual length $L<L_{\text{max}}$, the extra part is masked and not transmitted. At the BS, the received preamble is zero-padded to $L_{\text{max}}$. Since zero segments make feature importance unequal, we propose a preamble length adaptive module (PLAM). Inspired by \cite{ADJSCC}, PLAM concatenates the preamble length information $L$ with average-pooled input features, then uses a two-layer fully connected network to produce scaling factors. These adjust the Transformer input feature weights, enabling adaptive attention to cope with variable preamble lengths. It is worth noting that SA-UADNet-VPL captures task-oriented semantics by projecting pilots into a unified feature space and using self-attention together with PLAM to emphasize activity-related structures rather than full channel details, allowing consistent extraction of activity semantics across different pilot-length configurations.

To evaluate the performance of the proposed scheme, we used the probability of error in detecting user activity ($P_e$) as the performance metric. Fig. \ref{fig.aud}(b) presents $P_e$ as a function of preamble sequence length $L_{\text{test}}$. These results indicate that the proposed SA-UADNet-VPL scheme achieves superior error detection performance with shorter pilots compared to orthogonal matching pursuit (OMP) and approximate message passing (AMP) algorithms \cite{kml-tsp}, thereby effectively reducing access delay. Additionally, compared to models trained with fixed preamble lengths ($L_{\text{train}} = 10, 15, 18$), the scheme trained with variable-length preambles ($L_{\text{train}} \in [8,24]$) demonstrates enhanced generalization capabilities. Furthermore, by comparing SA-UADNet-VPL (with PLAM), $L_{\text{train}} \in \left[8,24\right]$, and SA-UADNet-VPL (no PLAM), $L_{\text{train}} \in \left[8,24\right]$, we can find that the inclusion of the PLAM further improves the detection performance, confirming the effectiveness of the proposed scheme in extracting user activity semantics from received signal.

\subsection{CSI Feedback}
\subsubsection{Related Work}
Traditional codebook or CS-based feedback requires high overhead, whereas deep learning reduces overhead and improves reconstruction. Most deep learning-based CSI feedback uses separate source-channel coding (SSCC), where encoder outputs are quantized into bits before channel coding. In SSCC frameworks, CSI reconstruction performance notably degrades when the signal-to-noise ratio (SNR) of the feedback link drops below a certain threshold, a phenomenon known as the cliff effect \cite{JSCC}. 
\begin{figure*}[tp!]
	\vspace*{-9mm}
	\begin{center}
		\hspace*{-4mm} \subfloat[]{
			\label{fig.csi_a}
			\includegraphics[width=0.66\textwidth,keepaspectratio]{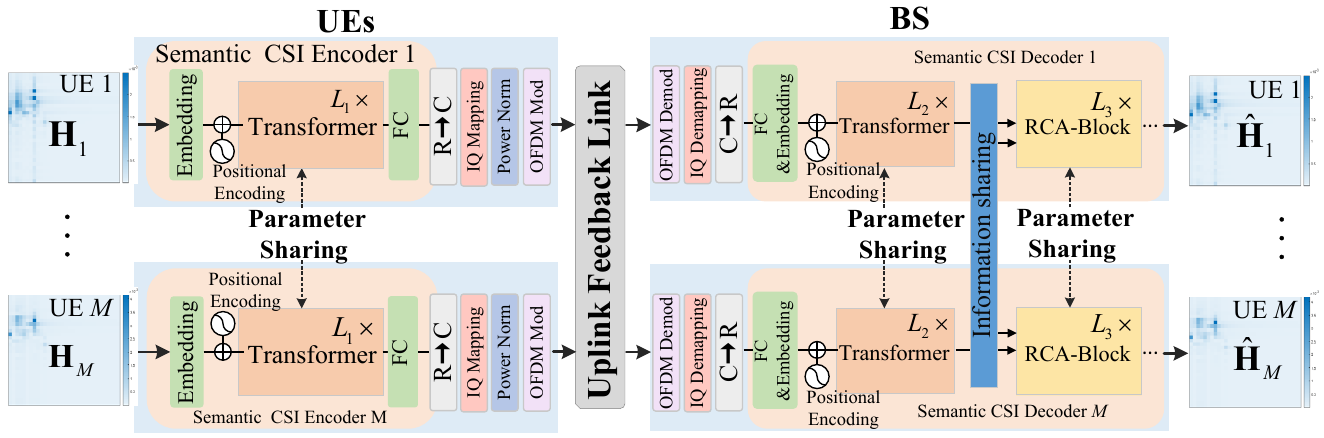}
		}
		\hspace*{-2mm}\subfloat[]{
			\label{fig.csi_c}
			\includegraphics[width=.31\textwidth,keepaspectratio]{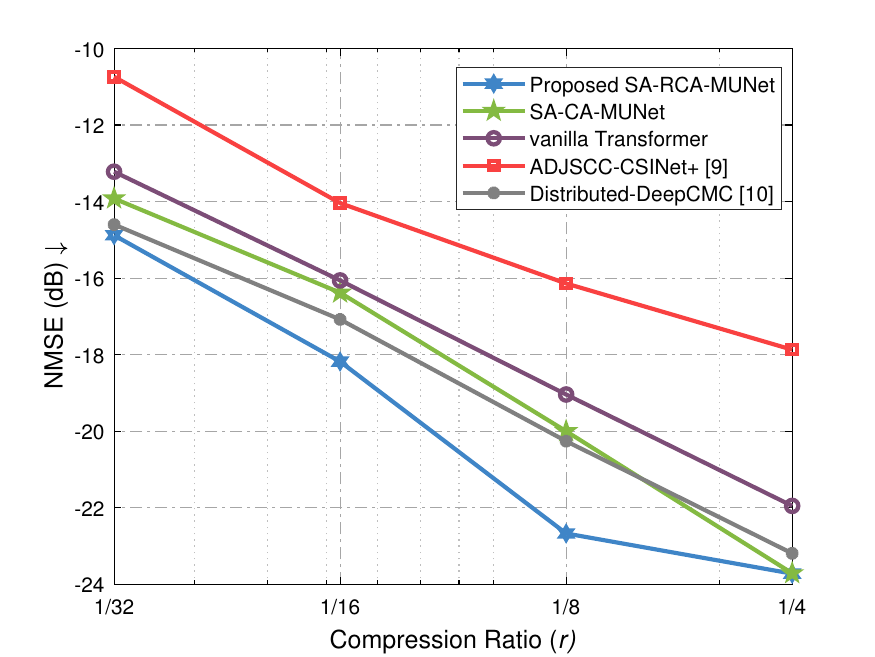}
		}
	\end{center}
	\vspace*{-5mm}
	\captionsetup{font={footnotesize}, name={Fig.},labelsep=period}
	\caption{(a)~The SA-RCA-MUNet-based multi-user semantic-aware CSI feedback architecture; and (b)~NMSE performance comparison of different CSI feedback schemes as a function of compression ratio\cite{zhw-wcl}. \textbf{Notes:} For performance evaluation, simulations used a dataset generated by Quadriga under the 3GPP TR 38.901 UMi scenario. The BS has a 32-antenna linear array, while the UE has a single antenna. The downlink channel operates at 6.7 GHz with 30 kHz subcarrier spacing across 1024 subcarriers. After angle-delay transformation, only the first $N_c=32$ delay-domain rows, where multipath energy concentrates, are fed back. The feedback uses $k$ subcarriers, yielding a compression ratio $r = k / (32 \times 32) = k/1024$.}
	\label{CSI-feedback} % Fig.4
	\vspace*{-4mm}
\end{figure*}
To address this, \cite{AnalogDeepCMC} proposed a convolutional neural network (CNN)-based deep joint source-channel coding (DJSCC) technique, using end-to-end optimized CNN encoders/decoders to overcome the cliff effect. \cite{CSI-JSCC} further proposed an attention-based JSCC (ADJSCC) scheme that integrates SNR information into encoder/decoder intermediate layers via an attention feature module \cite{ADJSCC}, enabling adaptation to varying SNRs. Considering the multi-user scenario, \cite{DeepCMC} proposed a distributed deep learning-based channel matrix compression (DeepCMC) scheme, which uses summation-based fusion in joint decoders to exploit geographically-correlated UEs’ side information. DeepCMC adopted a CNN-based structure, which  leaves potential for better capturing multi-user CSI semantic correlations.

\subsubsection{Semantic-Aware CSI Feedback}
We propose a semantic-aware residual cross-attention multi-user network (SA-RCA-MUNet) for CSI feedback, employing a Transformer-based backbone to utilize CSI semantics from geographically correlated UEs \cite{zhw-wcl}. As illustrated in the middle part of Fig. \ref{fig1}, 
each UE perceives and compresses downlink CSI into analog constellation symbols for feedback; the BS then enhances semantic CSI reconstruction using side information from other geographically correlated UE UEs.

In the multi-user CSI feedback system shown in Fig. \ref{CSI-feedback}(a), semantic features of delay-angle domain CSI for each UE are independently extracted and compressed using a Transformer-based semantic CSI encoder. After real-to-complex conversion, the outputs are mapped to analog constellation symbols and transmitted after in-phase/quadrature (IQ) mapping, power normalization, and orthogonal frequency division multiplexing (OFDM) mapping. At the BS, the received CSI undergoes OFDM demodulation, IQ demapping, and complex-to-real conversion, and is then reconstructed by the semantic CSI decoder.

Specifically, the Transformer acts as the backbone for both encoder and decoder. In the encoder, after converting complex to real numbers, the angle-delay domain CSI is represented by $N_t$ sequences of length $2N_c$, where $N_t$ is the number of BS antennas and $N_c$ the number of subcarriers. These sequences are embedded into $N_t$ feature sequences of length $d$, processed by an $L_1$-layer Transformer encoder for feature extraction, and finally compressed into a low-dimensional semantic sequence of dimension $2k$ via a linear layer for transmission.

Conversely, the decoder reconstructs the CSI by first applying dimensional upscaling via linear embedding, followed by a backbone with $L_2$ Transformer layers and $L_3$ residual cross-attention blocks (RCA-Blocks). The CSI for each UE is initially reconstructed through parameter-sharing Transformer layers, then refined jointly by RCA-Blocks. As detailed in \cite{zhw-wcl}, for each UE, the RCA uses query (Q) information from other UEs and key (K)/value (V) information from itself to compute attention scores and extract common information among different UEs. A residual cut is further introduced to provide complementary information across UEs. This leverages semantic channel information from geographically similar UEs and enhances CSI recovery.
In SA-RCA-MUNet, channel semantics are extracted via Transformer encoders and further refined through residual cross-attention, which selectively exchanges environment information among UEs. This injects complementary propagation semantics and leads to improved CSI reconstruction.
Besides, the normalized mean square error (NMSE) is chosen as the loss function for end-to-end optimization.

In Fig. \ref{CSI-feedback}(b), we compare the performance of various schemes at a received SNR of 10 dB under a two-UE scenario. Among these, SA-CA-MUNet replaces our residual cross-attention with typical cross-attention, while the vanilla Transformer simplifies SA-RCA-MUNet to a single-user network by substituting RCA-Blocks with conventional Transformers but maintaining the same number of network layers. ADJSCC-CSINet+ \cite{CSI-JSCC} is a CNN-based structure designed for single-user CSI feedback, and Distributed-DeepCMC extends the vanilla Transformer scheme to the multi-user case by incorporating the joint reconstruction method from \cite{DeepCMC}. These results demonstrate that significant gains in CSI reconstruction performance at various compression ratios can be achieved by perceiving and utilizing side information (i.e., channel semantic from nearby UEs) with the assistance of the proposed architecture.

\begin{figure*}[tp!]
	\vspace*{-8mm}
	\begin{center}
		\hspace*{-2mm}\subfloat[]{
			\label{fig.pre_a}
			\includegraphics[width=0.63\textwidth,keepaspectratio]{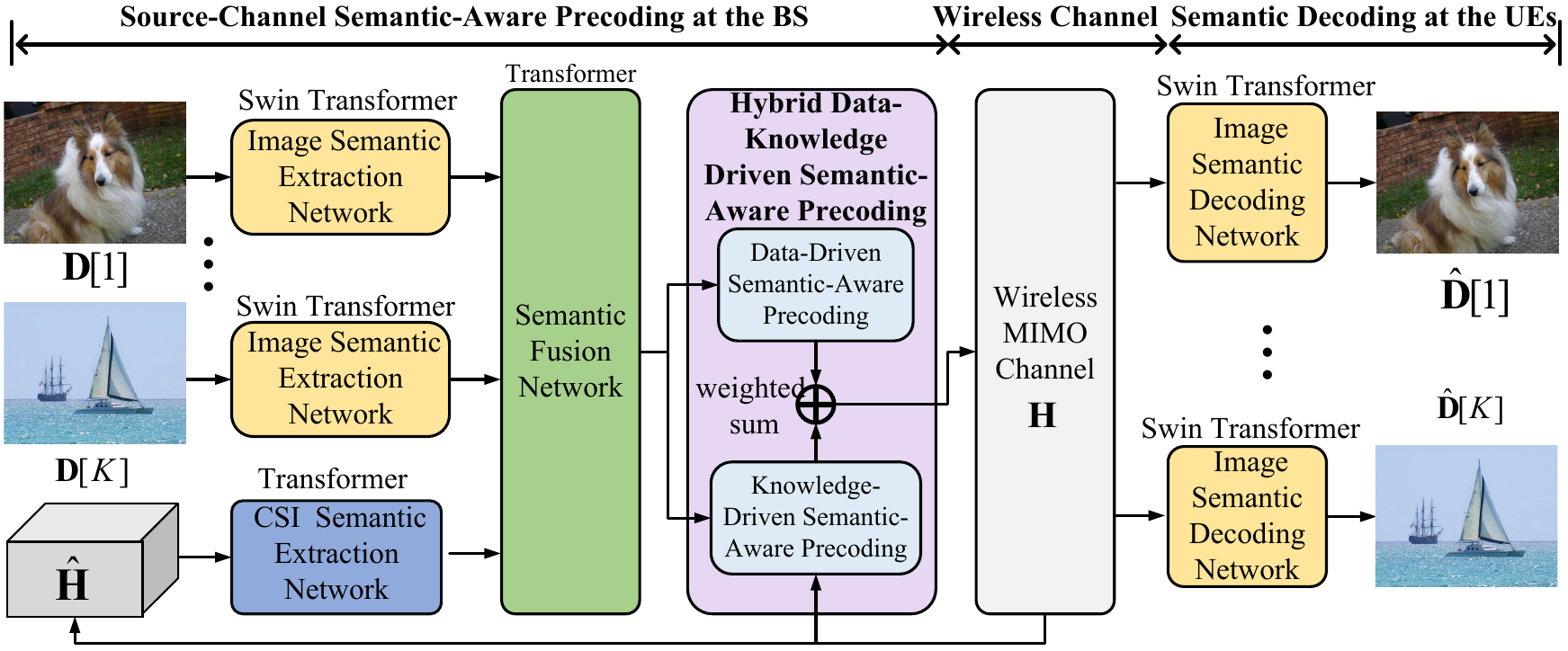}
		}
		\hspace*{-2mm}\subfloat[]{
			\label{fig.pre_b}
			\includegraphics[width=.36\textwidth,keepaspectratio]{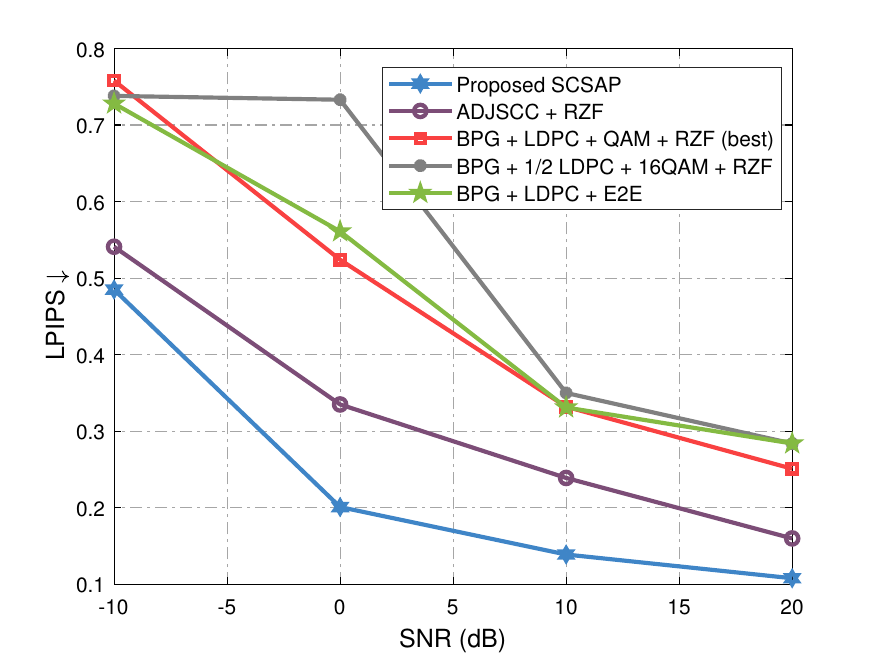}
		}
	\end{center}
	\vspace*{-5mm}
	\captionsetup{font={footnotesize}, name={Fig.},labelsep=period}
	\caption{(a)~Source-channel semantic-aware precoding for multi-user image transmission network; and (b)~LPIPS vs. SNR for different transmission schemes. \textbf{Notes:} It is assumed that a BS equipped with 64 antennas simultaneously transmits different image data to 4 UEs over the broadband multipath channel. Due to space limitations, we include only the key results here, while more complete analyses and full experimental settings can be found in \cite{WMH}.}
	\label{HBF} % Fig.5
	\vspace*{-4mm}
\end{figure*}
\subsection{Multi-user Downlink Hybrid Precoding}
\subsubsection{Related Work}
Multi-user hybrid precoding is essential for data transmission in massive MIMO systems. By accurately controlling signal directionality and mitigating user interference, precoding can substantially improve the system's spectral efficiency. Deep learning-based hybrid precoding methods have achieved lower complexity and better performance than traditional approaches. However, these methods mainly focus on optimizing spectral or energy efficiency under the assumption of ideal Gaussian-distributed sources, thus overlooking the semantic information of the source during precoding design.

To address both source semantics and CSI information in MIMO transmission, \cite{WYP} have proposed integrating CSI into the joint encoder-decoder framework. This integration of CSI and image semantic features facilitates adaptive power allocation according to feature importance, thereby improving the robustness of MIMO-based semantic transmission.

\subsubsection{Source-Channel Semantic-Aware Precoding (SCSAP)}
Despite these advances, existing research predominantly focuses on single UE transmission in small-scale MIMO systems. In contrast, we propose a joint design approach for massive MIMO systems, which fuses CSI and source semantics into the multi-user precoding design \cite{WMH}. As illustrated at the bottom of Fig. \ref{fig1}, the BS employs semantic CSI and data encoders to extract and fuse semantics for precoding, while each UE decodes its source information using a semantic data decoder.

More specifically,  as illustrated in Fig. \ref{HBF}(a), a Swin Transformer-based semantic extraction network processes high-dimensional image sources by leveraging window partitioning and shifting to reduce self-attention complexity while capturing multi-scale features. For the lower-dimensional CSI, a Transformer-based network extracts multi-user channel semantics. By concatenating the semantic features of the CSI and images at the feature dimension, a Transformer-based network is employed to integrate these semantic representations. This fusion strategy enables the subsequent precoder design to recognize both source and channel semantics, rather than relying solely on CSI as in conventional precoding designs. Besides, by embedding channel information directly into the source-semantic encoding process, the transmitted information can adapt to the underlying propagation conditions, thereby improving robustness in complex environments.

To enable efficient transmission of fused semantic features, we propose a hybrid data-knowledge driven semantic-aware precoding scheme \cite{WMH}. The data-driven semantic-aware precoding module feeds fused semantics into a Transformer, outputting precoded signals matching BS antenna dimensions. To improve robustness and interpretability, expert knowledge is integrated by designing a knowledge-driven semantic-aware precoding module, where a Transformer generates learnable parameters that replace traditional ones in the weighted minimum mean square error (WMMSE) algorithm. The precoding matrix obtained from learned WMMSE is then multiplied with the fused semantic features to generate the knowledge-driven precoded signals. 
Then, the outputs from both model-driven and knowledge-driven precoding are combined using a learnable weighted sum to produce the final precoded transmission signals. This method effectively leverages the strengths of expert knowledge and deep learning to optimize performance. Finally, each UE applies a Swin Transformer-based semantic decoder to reconstruct images from received feature maps. Unlike traditional precoding schemes focused only on spectral efficiency, the proposed SCSAP scheme uses semantic loss as its training criterion, aiming for high-quality image reconstruction. Moreover, SCSAP exploits image semantics by fusing image-derived source features with CSI-based propagation features, enabling a knowledge-driven precoder refined through a learnable semantic-correction branch. This hybrid design embeds source-distribution priors into the transmitter, allowing the precoder to better preserve perceptual semantics and surpass purely channel-driven precoding methods.

To evaluate the performance of the proposed SCSAP scheme, the learned perceptual image patch similarity (LPIPS) is adopted as the semantic performance metric and the following transmission schemes are compared: i) {ADJSCC + RZF}: Utilizes a CNN-based network architecture from \cite{CSI-JSCC} for joint source-channel coding.  To cope with the inter-user interference, regularized zero forcing (RZF) is used for multi-user precoding. ii) {BPG + LDPC + QAM + RZF (best)}: Through exhaustive search of different better portable graphics (BPG) source coding rates, low-density parity check (LDPC) channel coding rates, and quadrature amplitude modulation (QAM) orders, the combination with optimal  performance is selected. iii) {BPG + 1/2LDPC + 16QAM + RZF}: Uses fixed-rate source and channel coding and 16 QAM modulation. iv) {BPG + LDPC + E2E}: The system employs BPG for source coding and LDPC for channel coding. The single-user autoencoder-based end-to-end (E2E) transmission framework \cite{ofdm-e2e} is extended to accommodate the multi-user scenario for transmission. 

In Fig. \ref{HBF}(b), we compare the performance of above schemes at different received SNRs. The traditional SSCC method experienced significant performance degradation under low SNR conditions; in contrast, schemes based on joint source-channel coding design such as `ADJSCC + RZF' and the proposed `SCSAP' performed better. Furthermore, the proposed `SCSAP' scheme, which effectively leverages fused source-channel semantic information in precoding design, exhibits consistently better LPIPS performance over `ADJSCC + RZF', demonstrating the effectiveness of the semantic-aware precoding scheme. 

\section{Large Model-Enabled 6G Communications}\label{S3}
The previous section discussed specialist models for semantic-aware MIMO physical-layer communications. Recently, the general problem-solving capabilities demonstrated by LMs in fields such as natural language processing have led academia to explore their application in 6G. Beyond chat-based LMs for telecom domain knowledge, employing LMs for physical-layer tasks is also a noteworthy direction.

LMs typically undergo two important stages before practical application: pre-training and fine-tuning. Pre-training involves training model parameters on large-scale, non-specific task data to find a favorable initial point. Notably, the ``masked token prediction" paradigm used by the \textit{bidirectional encoder representations from transformers} (BERT) series and the ``next token prediction" adopted widely by the \textit{generative pretrained Transformer} (GPT) and \textit{large language model Meta AI} (LLAMA) series are deemed effective methodologies. Fine-tuning uses task-specific data to adjust pre-trained model parameters, enhancing performance on specific tasks, generally requiring much smaller datasets than pre-training. Unlike full parameter fine-tuning, parameter-efficient techniques such as \textit{low-rank adaptation} (LoRA), adapter tuning, and prefix tuning achieve comparable results by adjusting only a subset of the model's parameters, which also reduces training costs.

In Fig. \ref{fig5}, we evaluate wireless physical-layer works based on LMs by comparing different studies regarding model architecture, training methods, datasets used in pre-training, and fine-tuning methods along with applicable downstream tasks during the fine-tuning phase.

\subsection{Pre-training}
Existing research on pre-trained models can be classified into two primary approaches. The first approach employs \textit{text-based pre-trained models}, such as GPT-2 and LLAMA2, as key components for addressing communication tasks. This approach is often complemented by the fine-tuning of the backbone model and the input-output modules to adapt to downstream tasks. The second approach entails constructing a new foundation model pre-trained on channel datasets, i.e., \textit{CSI-based pre-trained models}. Similar fine-tuning techniques are then applied to tackle downstream tasks.

\subsubsection{Models Backbone}
\begin{figure*}[!t]
	%\vspace{-7mm}
	\centering
	\captionsetup{font={footnotesize}, name={Fig.},labelsep=period}
	\includegraphics[width=0.85\linewidth]{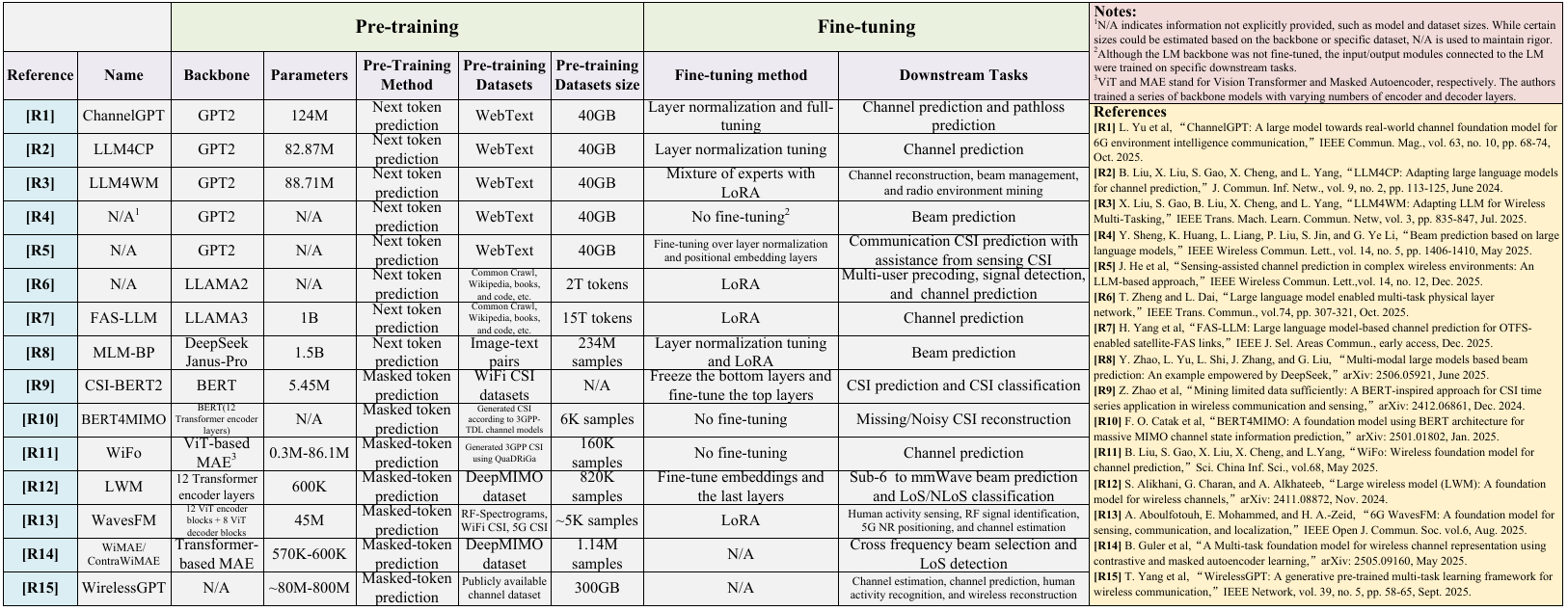}
	\caption{A comparison of various LM applications in communication physical-layer tasks.}
	\label{fig5}
	\vspace{-4mm}
\end{figure*}

In Fig. \ref{fig5}, studies [R1]-[R7] examine text-based pre-trained models, like GPT2 and LLAMA, which use Transformer decoder blocks and next-word prediction for pre-training. These LLMs have parameters ranging from millions (M) to billions (B).  Conversely, studies [R9]-[R13] utilize Transformer encoder blocks or masked autoencoder architectures for pre-training on CSI-type datasets, employing self-supervised learning methods like masked CSI token prediction. The parameter scale for CSI-based models varies significantly, from 300K to 800M, influenced by network layers and hidden space dimensions. Unlike the single-modal LMs discussed, [R8] explores a multimodal LM based on DeepSeek Janus-Pro, pre-trained on image and text data, to process both input types simultaneously.

\subsubsection{Datasets}
Text-based models for pre-training mainly use text-type datasets, such as Common Crawl, WebText, Wikipedia, books, and GitHub code. In contrast, CSI-based models rely on simulated or measured channel datasets like 3GPP channel models, DeepMIMO, and practical WiFi datasets. Some studies use diverse channel data types, like RF spectrograms, WiFi CSI, and 5G CSI [R13], or datasets from various geographical scenarios [R12, R14] to enhance pre-training comprehensiveness. Overall, dataset sizes for CSI-based models range from $5$K to $1$M samples. Compared to LLMs like LLAMA, with billions of parameters trained on trillions of tokens, CSI-based models face limitations due to the lack of large, publicly recognized datasets, posing a bottleneck in exploring larger communication models.

\subsubsection{Comparison Between Text/CSI Pretrained Models}

These two primary pre-training paradigms, text-pretrained LMs [R1]-[R8] and CSI-pretrained LMs [R9]-[R15], exhibit complementary strengths and inherent limitations.
Text-based LMs, though trained on linguistic data, use modality-independent Transformers capable of modeling long-range dependencies in arbitrary sequences. With appropriate tokenizers or projection heads, wireless inputs (pilots, CSI tensors, image semantic features) can be embedded into token spaces for Transformer processing, similar to multimodal LLMs (e.g., GPT-4V, LLaVA) using vision encoders for image patches. Additionally, parameter-efficient tuning (LoRA, adapters, MoE) further enables injecting communication-domain priors without modifying the full backbone. However, linguistic pretraining provides no inductive bias for physical-layer structures: standard tokenization ignores OFDM time–frequency correlation, antenna geometry, and channel statistics; generic LLMs also lack noise and interference awareness, risking misinterpretation of noise or reduced robustness under distribution shifts. Thus, while feasible, text-based LMs require domain adaptation for physical-layer tasks.
	
In contrast, CSI-pretrained models directly learn environment-related semantics but face data scarcity. Real-world CSI datasets (e.g., DeepMIMO) offer high fidelity but limited environmental and configuration diversity, while synthetic CSI from 3GPP, ray-tracing, or geometry-based models scales well yet exhibits a simulation-to-reality gap.
This can be addressed via generative modeling (GANs, diffusion) to generate diverse but physically consistent CSI, domain adaptation to align synthetic and real data, and wireless task-specific data augmentation (e.g., noise perturbation, dimension permutation).
Besides, in the long term, the release of extensive multi-environment CSI datasets by operators and research infrastructures represents an impactful approach to effective CSI-based pretraining.

\subsection{Fine-tuning}
Current fine-tuning methods typically fall into two main categories: 1) Adding additional input-output preprocessing modules for specific tasks without fine-tuning the backbone model parameters. For instance, in CSI prediction tasks, [R11] uses pre-trained channel models, and [R4] uses additional prompts with a trained input-output network for beam prediction.  2) Fine-tuning specific backbone model parameters along with input-output modules. Studies like [R3, R6, R7, R8, R13] apply LoRA by integrating low-rank matrices into the query and value matrices of the Transformer's attention module or feed-forward network. In contrast, [R1, R2, R5] adjust layer normalization parameters while freezing others. [R9, R12] explore fine-tuning the final layers of the backbone. Additionally, [R3] combines LoRA with MoE, using a gating network to select and combine experts for different tasks. This allows the network to leverage shared knowledge across tasks while retaining task-specific features for experts. Moreover, MoE uses sparse computation, activating a subset of experts per input, enabling model expansion for more tasks without increasing inference costs.

In downstream communication tasks, most research focuses on CSI-related tasks, including CSI prediction, CSI estimation, and classification tasks using CSI data, such as line of sight (LoS)/non-line of sight (NLoS) classification and human activity recognition. A small number of studies address tasks such as signal detection [R6], multi-user precoding [R6], and 5G new radio (NR) positioning [R13]. Additionally, some studies explore specific problem scenarios, for example, [R1], [R5], which examines the enhancement of CSI prediction through the incorporation of environment image or sensing CSI information. 

\subsection{Evolution From Specialist Models to Large Model}

Overall, the use of LMs in wireless communications remains in its early stage, and existing models in Section II mainly perform task-specific optimization. Although these specialists achieve strong performance, they depend on separate data pipelines and isolated representation spaces, limiting the reuse of common wireless priors such as uplink/downlink propagation consistency, spatial correlation, and noise/SNR statistics. This results in redundant parameters, duplicated training, and weak cross-task generalization. A unified LM can overcome these issues by employing a shared Transformer backbone with lightweight projection heads and task decoders [R3, R6, R15]. Heterogeneous inputs (e.g., received pilots, compressed CSI, image semantics) can be mapped into a unified token space, enabling scalable multi-task processing and cross-task knowledge transfer. In particular, recent CSI-based pre-trained backbones [R15] provide a promising direction: by learning latent environment structure, they extract reusable channel semantics for multiple tasks. Such unified representations highlight the potential of LMs to support higher-level reasoning over physical-layer signal structures.
	
Considering practical computational constraints, the evolution from specialist models to LMs can be conceptualized in three levels: \textit{(i) Level 1} is a BS-only LM that conducts multi-task inference using a shared backbone, facilitated by parameter-efficient methods like MoE for multiple tasks; \textit{(ii) Level 2} presents a hybrid design where UE-side specialist encoders generate semantic features that are aligned to the BS token space through projection heads; and \textit{(iii) Level 3} offers a complete BS-UE dual-side LM, where a compact UE backbone is derived through compression or distillation from the BS-side LM. The accompanying training strategy progresses naturally through these levels: Level 1 involves pretraining the BS backbone; Level 2 includes joint optimization of UE specialists with BS-side projection and task heads; and Level 3 entails end-to-end UL/DL fine-tuning for semantic alignment across the communication link. This provides a coherent and scalable path for evolving isolated specialist models into a unified foundation-model framework for wireless systems.

\section{Challenges and Open Issues}

In this section, we discuss potential future research directions in semantic-aware communication and LMs.

\textbf{Fundamental Theory of Communication LMs:}
LLMs follow an empirical scaling law indicating that test performance improves with increasing model size, data, and compute. However, classical signal detection and estimation tasks are constrained by limits such as the Cram\'{e}r-Rao lower bound and the Shannon-Hartley theorem, meaning that simply enlarging models or datasets may not yield uniform gains. Balancing model and data size is therefore essential for managing the complexity–-performance trade-off.
In higher-level semantic or pragmatic tasks, performance may exhibit patterns similar to LLM scaling behavior.
Moreover, large-model inference remains challenging for symbol-level physical-layer operations due to strict latency demands, though quantization, structured sparsification, and hardware-aware acceleration can meet frame-level latency for quasi-static tasks like UAD or CSI reconstruction. Therefore, developing latency-efficient communication LMs is thus crucial for improving real-time applicability.
	
\textbf{Multimodal Channel/Environment Information Driven LMs:}
Existing CSI-based LMs are mostly pre-trained on 3D CSI tied to specific frequency bands and scenarios, relying on a single modality and insufficiently using cross-band, cross-scenario, and cross-modal data such as radio maps, visual inputs, and radar signals. Consequently, leveraging multimodal channel/environment information to train LMs that capture richer features for tasks like beam prediction, radio map reconstruction, and sensing remains a major challenge for 6G semantic-aware MIMO. Notably, \cite{DS} studied the multimodal LM DeepSeek Janus-Pro-1B, using its cross-modal alignment to fuse image data with UE positions for improved beam prediction, showing strong potential for communication applications. Further exploration is also needed on pre-training and fine-tuning strategies for such multimodal LMs.

\textbf{Generative Models-Driven Semantic-Aware Communications:}
Semantic communication targets transmitting and interpreting diverse sources such as text, speech, images, and CSI. Generative models can create new samples following desired conditional distributions by learning from large datasets. Thus, by sending only essential semantic information and allowing a generative model (e.g., diffusion models or LMs) to reconstruct content at the receiver using source-distribution priors, communication overhead can be greatly reduced. For example, in image transmission, \cite{QL} proposes sending only a ``textual prompt + edge map'' as the semantic representation, enabling a conditional diffusion model at the receiver to reconstruct a high-fidelity image with substantially lower cost and latency.

\section{Conclusions}
This article has presented the paradigm evolution of semantic-aware MIMO communication. Through the examination of tasks such as user activity detection, CSI feedback, and precoding, we have demonstrated that specialist models can effectively perceive, utilize, and fuse the semantic features of sources and channels, thereby enhancing MIMO communication performance. Additionally, the study has explored the role of large models in various physical-layer tasks, revealing the potential for enabling intelligent networks. Finally, we have identified unresolved challenges within large communication models and discussed future research directions toward 6G semantic-aware MIMO communications.

\bibliographystyle{IEEEtran}

\end{document}